\documentclass[12pt]{article} 
\renewcommand{\baselinestretch}{1.5}
\usepackage[dvipsnames]{xcolor}
\usepackage{graphicx}
\usepackage{amsmath}
\usepackage{amssymb} 
\usepackage{slashed}
\usepackage{bm}
\usepackage{here}
\usepackage{cite}
\usepackage{ulem}

\def\ignore#1{{}}

\def\SM{{\rm SM}}
\def\NP{{\rm NP}}
\def\KK{{\rm KK}}

\def\mynoalign{\noalign{\kern 4pt}}
\def\mysnoalign{\noalign{\kern 3pt}}

\newcommand{\LR}{\text{LR}}
\newcommand{\GHU}{\text{GHU}}

\DeclareMathOperator{\diag}{diag}
\begin{document}

\thispagestyle{empty}


\baselineskip=16pt
\leftline{June 18, 2022\hfill OU-HET-1108}
\rightline{KYUSHU-HET-237}

\vskip 2.5cm

\baselineskip=30pt plus 1pt minus 1pt
\begin{center}
{\LARGE \bf Bhabha scattering}
\\
{\LARGE \bf in the gauge-Higgs unification}
\end{center}
%
%
\baselineskip=22pt plus 1pt minus 1pt
\vskip 1.5cm
\begin{center}
{\bf Shuichiro Funatsu$^1$, Hisaki Hatanaka$^2$, Yutaka Hosotani$^3$,}
{\bf Yuta Orikasa$^4$ and Naoki Yamatsu$^5$}
\baselineskip=17pt plus 1pt minus 1pt
\vskip 10pt
{\small \it $^1$Institute for Promotion of Higher Education, Kobe University, Kobe 657-0011, Japan} \\
{\small \it $^2$Osaka, Osaka 536-0014, Japan} \\
{\small \it $^3$Department of Physics, Osaka University, 
Toyonaka, Osaka 560-0043, Japan} \\
{\small \it $^4$Institute of Experimental and Applied Physics, Czech Technical University in Prague,} \\
{\small \it Husova 240/5, 110 00 Prague 1, Czech Republic} \\
{\small \it $^5$Department of Physics, Kyushu University, Fukuoka 819-0395, Japan} \\
\end{center}
\vskip 1.5cm
\baselineskip=16pt plus 1pt minus 1pt
\begin{abstract}
We examine effects of $Z'$ bosons
in  gauge-Higgs unification (GHU) models at  
$\mathrm{e}^{+}\mathrm{e}^{-}\to \mathrm{e}^{+}\mathrm{e}^{-}$ Bhabha scatterings.
We evaluate  differential cross sections in Bhabha scatterings including  $Z'$ bosons 
in two types of $\mathrm{SO}(5) \times \mathrm{U}(1) \times \mathrm{SU}(3)$ GHU models.
We find that deviations of differential cross sections in the GHU models from those in the SM can be 
seen at $\sqrt{s} = 250\,\mathrm{GeV}$.
With $80\,\%$-longitudinally polarized electron and $30\,\%$-longitudinally polarized positron beams, 
the left-right asymmetries in the GHU A- and B-models are  resolved 
at more than $3\,\sigma$ at $L_{\rm int} =250\,\mathrm{fb}^{-1}$.
We also show that Bhabha scattering  with scattering angle less than 100 mrad can be safely used 
as  luminosity measurements at $\mathrm{e}^{+}\mathrm{e}^{-}$ colliders since the effects of  $Z'$ bosons 
are well suppressed for small scattering angle.
We propose a new observable which can be measured at future TeV-scale $\mathrm{e}^{+} \mathrm{e}^{-}$ linear colliders.
\end{abstract}
%
%
%
\newpage
\baselineskip=20pt plus 1pt minus 1pt
\parskip=0pt
\newcommand{\calM}{\mathcal{M}}
\newcommand{\calO}{\mathcal{O}}
\renewcommand{\Re}{\mathrm{Re}}
\newcommand{\rme}{\mathrm{e}}
\newcommand{\SO}{\mathrm{SO}}
\newcommand{\SU}{\mathrm{SU}}
\newcommand{\U}{\mathrm{U}}
\newcommand{\AdS}{\mathrm{AdS}}
\newcommand{\GeV}{\text{GeV}}
\newcommand{\TeV}{\text{TeV}}
\newcommand{\Lint}{L_{\mathrm{int}}}
\newcommand{\fbi}{\mathrm{fb}^{-1}}
\newcommand{\abi}{\mathrm{ab}^{-1}}
\section{Introduction}

With the establishment of the standard model (SM) by the discovery of the Higgs boson, searching for physics 
beyond the SM (BSM) and understanding the electroweak phase transition have become main topics in  particle physics.
As one scenario of BSM, the gauge-Higgs unification (GHU) scenario has been studied
\cite{Manton,Hosotani1983,Hosotani1988,
Davies:1987,Davies:1988,
HIL,Hatanaka1999,Kubo:2001zc,Burdman:2002se,
Csaki:2002,
Scrucca2003,
Agashe,
Cacciapaglia2006,
Medina,
Falkowski}.
In GHU, the Higgs boson is a part of the extra-dimensional 
component of gauge potentials, appearing as a fluctuation mode of an 
Aharonov-Bohm (AB) phase $\theta_H$ in the fifth dimension.
Many GHU models are proposed for electroweak unification \cite{
Hosotani:2007qw,
Hosotani:2008by,
Funatsu:2013ni,
Funatsu:2014fda,
Funatsu:2016uvi,
Funatsu:2017nfm,
Hatanaka:2013iya,
Funatsu:2019xwr,
Funatsu:2019fry,
Funatsu:2020haj,
Funatsu2019a,
Funatsu-FT,
Funatsu:2021yrh,
Yoon2017,Yoon2018,Yoon2019,
Kurahashi:2014jca,
Matsumoto:2016okl,
Hasegawa:2018jze,
Kakizaki:2021kof
},
and  GHU models for grand unification are also proposed \cite{
Lim2007,
Hosotani:2015hoa,
Furui,
Hosotani:2017edv,
Englert:2019xhz,
Englert:2020eep,
Kakizaki:2013eba,
Kojima:2017qbt,
Maru:2019bjr,
Angelescu:2021nbp
}. 
Among them, two types of $\SU(3)_{C} \times \SO(5) \times \U(1)_{X}$ GHU models,
the A-model\cite{
Hosotani:2007qw,
Hosotani:2008by,
Funatsu:2013ni,
Funatsu:2014fda,
Funatsu:2016uvi,
Funatsu:2017nfm,
Hatanaka:2013iya}
and B-model\cite{Funatsu:2019xwr,Funatsu:2019fry,Funatsu:2020haj,
Funatsu2019a,
Funatsu-FT,
Funatsu:2021yrh},  in warped spacetime have been extensively studied. 
At low energies below the electroweak scale the mass spectrum and gauge and Higgs couplings of
SM particles are nearly the same as in the SM.

Couplings of the first Kaluza-Klein (KK) neutral gauge bosons to right-handed SM fermions are large in the A-model,
whereas those to left-handed SM fermions become large in the B-model.
In proton-proton collisions KK bosons of photons and $Z$ bosons appear as huge broad resonances
of $Z'$ bosons in the Drell-Yan process, and can be seen in current and future hadron collider experiments
 \cite{Funatsu:2014fda,Funatsu:2016uvi,Funatsu:2021yrh}.
The KK-excited states of the $W$ boson are also seen as resonances of $W'$ bosons. 
In the A-model the couplings of the first KK $W$ boson to 
the SM fermions are small.
In the B-model the couplings of the first KK $W$ boson to right-handed fermions are negligible, 
while the couplings to left-handed fermions are much larger than the $W$ boson couplings.
Therefore at the LHC the first KK $W$ boson appears as a  narrow resonance of $W'$ boson in the A-model, 
but appears as a broad resonance in the B-model\cite{Funatsu:2016uvi,Funatsu:2021yrh}.

In  $\rme^{+}\rme^{-}$ collider experiments,   effects of GHU can be examined
by exploring interference effects among photon, $Z$ boson and $Z'$ bosons.
In the previous papers we have studied  effects of new physics (NP) on
such observable quantities  as cross section, forward-backward asymmetry and left-right asymmetry 
in $\rme^{+}\rme^{-} \to f\bar{f}$ ($f\ne \rme$) with polarized and unpolarized $\rme^{+}\rme^{-}$ beams
\cite{Funatsu:2017nfm,Funatsu:2020haj,
Yoon2018,
Yoon2019,
Bilokin2017,
Irles2019,
Irles2020}.
In Ref.\ \cite{Funatsu:2017nfm} we compared such observable quantities of GHU with those of the SM in LEP experiments
at $\sqrt{s} = M_Z$, and LEP2 experiments for $130 \GeV \le \sqrt{s} \le 207 \,\GeV$
\cite{ALEPH:2005ab,ALEPH:2013dgf}.
In Refs.~\cite{Funatsu:2017nfm,Funatsu:2020haj} we also gave predictions of several  signals of $Z'$ bosons in 
GHU in  $\rme^{+}\rme^{-} $ collider experiments 
designed  for future  with collision energies $\sqrt{s} \ge 250 \text{ GeV}$ with polarized electron and positron beams.
In the $\rme^{+}\rme^{-} \to f\bar{f}$ ($f\ne \rme$) modes,
the deviations of total cross sections become large for
right-polarized electrons in the A-model, whereas
in the B-model the deviations are large for left-polarized electrons.

Deviations from the SM can be seen in the Higgs couplings as well.
$HWW$, $HZZ$ and Yukawa couplings deviate from those in the SM in a universal manner\cite{Hosotani:2007qw,Hosotani:2008by,Kurahashi:2014jca}.
They are suppressed by a common factor 
\begin{eqnarray}
\frac{g_{HWW}^{\GHU}}{g_{HWW}^{\SM}},\,
\frac{g_{HZZ}^{\rm \GHU}}{g_{HZZ}^{\SM}}
\simeq
\cos(\theta_H) 
\end{eqnarray}
for $W$ and $Z$ bosons, and 
\begin{eqnarray}
\frac{y_{\bar{f}f}^{\GHU}}{y_{\bar{f}f}^{\SM}}
\simeq
\begin{cases}
\cos(\theta_{H}) & \text{A-model \cite{Hosotani:2008by}} \\
\cos^{2}(\theta_{H}/2) & \text{B-model \cite{Funatsu:2019fry}} 
\end{cases}
\label{Hcoupling1}
\end{eqnarray}
for SM fermions $f$.
In the analysis of both $Z'$ and $W'$ bosons in hadron colliders
\cite{Funatsu:2014fda,Funatsu:2016uvi,
Funatsu:2021yrh},
it is found that the AB phase is constrained as $\theta_{H} \lesssim 0.1$.
For $\theta_H = \calO(0.1)$,  probable values in the model,    
the deviation of the couplings amounts to $(1-\cos\theta_H) = \calO(0.005)$, and is small.
At the International Linear Collider (ILC) at $\sqrt{s} = 250\GeV$, the $ZZH$ coupling can be measured 
in the $0.6\,\%$ accuracy with $2\text{ ab}^{-1}$ data \cite{Barklow:2017suo}.
Since the masses and Higgs couplings of the SM fields in the GHU models are very close to those in the SM, 
the electroweak phase transition (EWPT) occurs
at $T_{C} \sim 160 \GeV$ and appears very weak first order\cite{Hatanaka:2013iya,Funatsu-FT}
in both A- and B-models, which is very similar to EWPT in the SM\cite{Senaha:2020mop}.

In this paper we study  effects of $Z'$ bosons in GHU models
on the $\rme^{+} \rme^{-} \to \rme^{+} \rme^{-}$ Bhabha scattering.
Measurements of  Bhabha scattering at  linear colliders have
contributed to the establishment of the SM\cite{
Abe:1994sh,
Abe:1994wx,
Abe:1996nj
}.
Bhabha scattering is also useful to explore NP \cite{
Richard:2018zhl,
Das:2021esm
}.
Unlike $\rme^{+} \rme^{-} \to f\bar{f}$ ($f\ne \rme$) scattering, 
in Bhabha scattering not only $s$-channel but also  $t$-channel contributions enter the process.
Since the $t$-channel contribution of photon exchange dominates in   forward scattering amplitudes,
the cross section becomes very large for  small scattering angles,
which improves the statistics of experiments.
It will be seen below that effects of $Z'$ bosons on cross sections can be measured 
with large significances.

Bhabha scattering at very small scattering angles
is used for the determination of the luminosity of the $\rme^{+}\rme^{-}$ beams. 
Since cross sections of  all other scattering processes depend on the luminosity, 
one needs to know whether or not effects of $Z'$ bosons on the $\rme^{+}\rme^{-} \to \rme^{+}\rme^{-}$ cross section
are sufficiently small.
Forward-backward asymmetry $A_{FB}$ of the cross section in Bhabha scattering  is no longer a good quantity 
for searching NP, since the backward scattering cross section is much smaller than the forward scattering cross section.
We will propose a new quantity $A_X$ to measure with polarized $\rme^{+}\rme^{-}$ beams,
which can be used for seeing NP effects instead of $A_{FB}$.

In Section 2 we briefly review the GHU A- and B-models  and 
discuss the $\rme^{+}\rme^{-} \to \rme^{+}\rme^{-}$ scattering 
in both the SM and GHU models. In Section 3, we show the formulas of
$\rme^{+}\rme^{-}\to \rme^{+}\rme^{-}$ scattering cross sections for longitudinally polarized $\rme^{\pm}$ beams, 
and numerically evaluate the effects
of $Z'$ bosons in GHU models on  differential cross sections and left-right asymmetries.
We also show that effects of $Z'$ bosons on the cross section
are very small at the very small scattering angle.

\section{Gauge-Higgs unification}

In GHU A- and B-models the electroweak $\SU(2)\times\U(1)$ symmetry
is embedded in $\SO(5)\times\U(1)_{X}$ symmetry in the Randall-Sundrum warped space \cite{Randall:1999ee},
whose metric is given by
\begin{align}
ds^{2} &= \frac{1}{z^{2}} \left[\eta_{\mu\nu}dx^{\mu}dx^{\nu} 
+ \frac{dz^{2}}{k^{2}}
\right],
\quad
1 \le z \le z_{L} = e^{kL}
\end{align}
where $\eta_{\mu\nu} = \diag(-1,+1,+1,+1)$ and $k$ is the AdS-curvature.
 We refer two 4D hyperplanes at $z=1$ and $z=z_{L}$ as the UV and IR branes, respectively.
The $\SO(5)$ symmetry is broken to $\SO(4)\simeq \SU(2)_{L} \times \SU(2)_{R}$ by the boundary conditions at $z=1,\,z_{L}$ and
the $\SU(2)_{R} \times \U(1)_{X}$ symmetry is broken to $\U(1)_{Y}$ by a scalar field 
localized on the UV brane.
$\SU(2)_{L}\times\U(1)_{Y}$ symmetry is broken to the 
electromagnetic $\U(1)_{\rm EM}$ symmetry by
the VEV of the $z$-component gauge fields of $\SO(5)/\SO(4)$. 
The VEV is related to the gauge-invariant AB phase $\theta_{H}$. 
%
%
\begin{table}[tbh]
    \renewcommand{\arraystretch}{1.2}
    \centering
     \caption{The fermion content in the first generation of the lepton sector is shown.
       In the A-model bulk fermions are introduced in the vector representation of $\SO(5)$, which are decomposed of 
       $\SO(4)$-vector and singlet. Zero modes of fermions appear in the left-handed components of $\SO(4)$-vectors 
       and in the right-handed components of $\SO(4)$-singlets.
     The extra zero modes of $L_{j}$ ($j=1,2,3$) couple to brane fermions at the UV brane 
     to become massive\cite{Funatsu:2013ni}. In the B-model bulk fermions are introduced in the spinor representation 
     of $\SO(5)$. Zero modes appear in the left-handed components of $\SU(2)_L$ doublet $(\nu_{\rme},\rme) $ and  
     in the right-handed components of $\SU(2)_R$ doublet $(\nu'_{\rme}, \rme')$ \cite{Funatsu:2019xwr}.
       }
    \vskip 10pt
    \begin{tabular}{|c|c|c|}
    \hline
    & A-model & B-model \\
    \hline \hline
    Bulk fermion
    &$ \begin{matrix}
    \left(
    \begin{pmatrix} \nu_{\rme} & L_{1X} \\ \rme & L_{1Y}
    \end{pmatrix},\, \rme'\right),
    \\
    \left(
    \begin{pmatrix}
     L_{2X} & L_{3X} \\ L_{2Y} & L_{3Y}
    \end{pmatrix},\,
    \nu'_{\rme}
     \right)
    \end{matrix}
    $
    &
    $
    \begin{pmatrix} \nu_{\rme} \\  \rme \\ \nu'_{\rme} \\ \rme' \end{pmatrix}
    $ \rule[-3mm]{0mm}{8mm}
    \\
    \hline
    \end{tabular}
    \label{Tab:matter}
\end{table}
 
The difference between the A- and B-models lies in the content of fermions
as tabulated in Table~\ref{Tab:matter}.
In the A-model, quarks and leptons in the SM are embedded in the $\SO(5)$-vector representation. 
In the B-model quarks and leptons are embedded in the $\SO(5)$-spinor, vector and singlet representations, which are naturally derived from spinor and vector multiplets in the $\SO(11)$ 
gauge-Higgs grand unification\cite{Hosotani:2015hoa, Furui}. 
We also note that the bulk mass parameter for the bulk electron field, $c_{\rme}$, is positive in
the A-model \cite{Funatsu:2013ni} whereas $c_{\rme}$ 
in the B-model has to be negative\cite{Funatsu:2019xwr}. In the B-model   positive $c_{\rme}$ leads to
an exotic light neutrino state and therefore negative $c_{\rme}$ must be chosen.
Zero modes of fermion fields with positive bulk mass parameters are localized near the UV brane
whereas zero modes of fermion fields with negative bulk mass parameters are localized near the IR brane.

Interactions of the electron and gauge bosons are given by
\begin{align}
\int d^{4}x \int_{1}^{z_{L}} \frac{dz}{k}
\biggl\{
\bar{\check{\Psi}} \gamma^{\mu} (\partial_{\mu} - i g A_{\mu}) \check{\Psi}
\biggr\}
\end{align}
where $A_{\mu}(x,z)$ is a four-dimensional component of the 5D gauge field and
$\Psi(x,z) = z^2 \check{\Psi}$ is the 5D electron field.
The electron corresponds to the zero mode of $\check{\Psi}(x,z)$.
In the A-model  the left-handed electron is localized in the vicinity of the UV brane and the right-handed 
component is localized near the IR brane.
In the B-model the right-handed electron is localized in the vicinity of the UV brane and the left-handed 
component is localized near the IR brane.
$A_{\mu}(x,z)$ has a KK expansion which contains the photon, $Z$ boson and their KK excited modes. 
The wave function of the photon is constant in the fifth dimension coordinate $z$.
The wave function of  the $Z$ boson is almost constant in $z$, but has nontrivial dependence near the IR brane. 
Couplings of the electron to the $Z$ boson are very close to those in the SM.
Wave functions of the first KK-excited modes of gauge bosons  are localized near the IR brane so that 
the first KK-excited gauge bosons  couple  strongly with   fermions localized near the IR brane.
In the A-model  right-handed electrons have large couplings to the first KK-excited gauge bosons 
whereas in the B-model left-handed electrons couple strongly to the first KK-excited gauge bosons. 

In Tables~\ref{tbl:model} and \ref{tbl:couplings}, parameters and couplings in the A- and B-models are tabulated.
Here, model parameters ($\theta_H,m_{\KK}$ and $z_{L}$),
masses, widths and couplings of $Z'$-bosons
are referred from Refs.~\cite{Funatsu2019a,Funatsu:2020haj}.
\begin{table}[t]
\caption{Parameters in GHU models. $c_{\rme}$ in the rightest column is
 the bulk parameter for the electron field.}\label{tbl:model}
\vspace{5mm}
\centering
\begin{tabular}{c|ccc|ccccccc}
\hline\hline
Model & $\theta_{H}$ & $m_{\KK}$ & $z_{L}$ 
& $m_{\gamma^{(1)}}$ & $\Gamma_{\gamma^{(1)}}$  
& $m_{Z^{(1)}}$ & $\Gamma_{Z^{(1)}}$  
& $m_{Z_{R}^{(1)}}$ & $\Gamma_{Z_{R}^{(1)}}$ & $c_{\rme}$  
\\
& [rad] & [TeV] & & [TeV] & [TeV] & [TeV] & [TeV] & [TeV] & [TeV] &
\\
\hline
A & $0.08$ & $9.54$ & $1.01\times10^{4}$
& $7.86$ & $0.99$
& $7.86$ & $0.53$
& $7.31$ & $1.01$ & 2.0342
\\
B & $0.10$ & $13.0$ & $3.87\times10^{11}$ 
& $10.2$ & $3.25$ 
& $10.2$ & $7.84$ 
& $9.95$ & $0.816$ & $-1.0067$ 
\\
\hline
\end{tabular}
\end{table}
The big difference in the magnitude  of $z_{L}$  in the A- and B-models
originates from the formulas of top-quark mass.
In the A-model 
$
m_{\rm top}^{\rm A} \simeq (m_{\KK}/(\sqrt{2}\pi)) \sqrt{1 - 4c_{\rm top}^{2}} \sin\theta_{H}
$ \cite{Funatsu:2013ni}
whereas in the B-model 
$
m_{\rm top}^{\rm B} \simeq (m_{\KK}/\pi) \sqrt{1-4c_{\rm top}^{2}} 
\sin\frac{1}{2}\theta_{H}$ \cite{Funatsu:2019xwr}.
In both models the $W$ boson mass is given by
$m_{W} \simeq  m_{\KK}/(\pi \sqrt{kL}) \sin\theta_{H}$
so that the lower bound of $z_{L}$ becomes
$z_{L} \gtrsim 8\times10^{3}$ in the A-model  and $z_{L} \gtrsim 7 \times 10^{7}$ in the B-model.
In Table~\ref{tbl:model} the bulk mass parameter for the electron field $c_{\rme}$ is given.
As explained before, $c_{\rme}$ is positive in the A-model whereas 
$c_{\rme}$ is negative in the B-model.
\begin{table}[t]
\caption{
Left-handed and right-handed couplings of the electron, 
$\ell_{V},r_{V}$ ($V = Z,Z^{(1)}$, $Z_{R}^{(1)}$ and $\gamma^{(1)}$), in units  of $g_{w}\equiv g_{A}/\sqrt{L}$ (see text).
Ratios of $e$ and $g_{w}$ are shown in the second column.}\label{tbl:couplings}
\vspace{5mm}
\centering\small
\begin{tabular}{c|cccccccccc}
\hline\hline
Model & $(e/g_{w})^{2}$ 
 & $\ell_{Z}$ & $r_{Z}$ 
 & $\ell_{Z^{(1)}}$ & $r_{Z^{(1)}}$ 
 & $\ell_{Z_{R}^{(1)}}$ & $r_{Z_{R}^{(1)}}$ 
 & $\ell_{\gamma^{(1)}}$ & $r_{\gamma^{(1)}}$ & 
\\
\hline
A & $0.2312$
& $-0.3066$ & $0.2638$
& $0.1195$ & $0.9986$
& $0.0000$ & $-1.3762$ 
& $0.1879$ & $-1.8171$ \\
B  & $0.2306$ 
& $-0.3058$ & $0.2629$ 
& $-1.7621$ & $-0.0584$ 
& $-1.0444$ & $0.0000$ 
& $-2.7587$ & $0.1071$
\\
\hline
\end{tabular}
\end{table}
In Table~\ref{tbl:couplings}, the left- and right-handed electron couplings to $Z'$ bosons, $r_{V}, \ell_{V}$ ($V = Z,Z^{(1)},Z_{R}^{(1)},\gamma^{(1)}$), are tabulated.
In the table $g_{w}\equiv g_{A}/\sqrt{L}$ is the 4D gauge coupling of the $\SO(5)$ where
$g_{A}$ is the 5D $\SO(5)$ coupling.
In terms of $g_{A}$ and the 5D $\U(1)_{X}$ coupling $g_{B}$, a mixing parameter is defined as \cite{Funatsu:2014fda,Funatsu:2020haj}
\begin{align}
e/g_{w} = \sin\theta_{W}^{0} &\equiv 
\frac{s_{\phi}}{\sqrt{1 + s_{\phi}^{2}}},
\quad
s_{\phi} \equiv g_{B}/\sqrt{ g_{A}^{2} + g_{B}^{2}}.
\label{bareWangle}
\end{align}
The value of $\sin\theta_{W}^{0}$ is determined so as to reproduce 
the experimental value of the forward-backward asymmetry in 
$\rme^{+}\rme^{-}\to \mu^{+}\mu^{-}$ scattering at the $Z$-pole.
In the A-model electron's right-handed couplings to $Z'$-bosons are larger than left-handed couplings.
In the B-model electron's left-handed couplings to $Z'$-bosons are larger than right-handed couplings.

\section{Bhabha scattering in $\rme^{+}\rme^{-}$ colliders}

We consider the $\rme^{+} \rme^{-} \to \rme^{+} \rme^{-}$ scattering in the center-of-mass frame.
In this frame, the Mandelstam variables $(s,t,u)$ are given by
\begin{align}
s &= 4 E^{2},
\nonumber \\
t &=-\frac{s}{2}(1-\cos\theta) = -s \sin^{2}\frac{\theta}{2},
\nonumber \\
u &=-\frac{s}{2}(1+\cos\theta) = -s \cos^{2} \frac{\theta}{2}
\end{align}
where $E$  is the energy of initial electron and positron, and $\theta$ is the scattering angle of the electron.
Since the $\rme^{+}\rme^{-} \to \rme^{+}\rme^{-}$ scattering process consists
both $s-$ and $t-$channel processes,
the scattering amplitude is written in terms of the following six building blocks:
\begin{align}
S_{LL} = S_{LL}(s) 
&\equiv 
\sum_{i} \frac{\ell_{V_{i}}^{2}}{s - M_{V_{i}}^{2} + iM_{V_{i}}\Gamma_{V_{i}}},
\nonumber \\
S_{RR}= S_{RR}(s) 
&\equiv
\sum_{i} \frac{r_{V_{i}}^{2}}{s - M_{V_{i}}^{2} + iM_{V_{i}}\Gamma_{V_{i}}},
\nonumber \\
S_{LR} = S_{LR}(s) &\equiv
\sum_{i} \frac{\ell_{V_{i}} r_{V_{i}}}{s - M_{V_{i}}^{2} + iM_{V_{i}}\Gamma_{V_{i}}},
\nonumber \\
T_{LL} = T_{LL}(s,\theta) &\equiv
\sum_{i} \frac{\ell_{V_{i}}^{2}}{t - M_{V_{i}}^{2} + iM_{V_{i}}\Gamma_{V_{i}}},
\nonumber \\
T_{RR} = T_{RR}(s,\theta) &\equiv
\sum_{i} \frac{r_{V_{i}}^{2}}{t - M_{V_{i}}^{2} + iM_{V_{i}}\Gamma_{V_{i}}},
\nonumber \\
T_{LR} = T_{LR}(s,\theta) &\equiv 
\sum_{i} \frac{\ell_{V_{i}} r_{V_{i}} }{t - M_{V_{i}}^{2} + iM_{V_{i}}\Gamma_{V_{i}}},
\label{eq:blocks}
\end{align}
where $M_{V_{i}}$ and $\Gamma_{V_{i}}$ are the mass and width of the vector boson $V_{i}$.
$\ell_{V_{i}}$ and $r_{V_{i}}$ are left- and right-handed couplings of electrons to the vector boson 
$V_{i}$ ($V_{0} =\gamma$, $V_{1} = Z$), respectively. 
In particular, we have
$\ell_{\gamma} = r_{\gamma} = Q_{\rme} e$, $Q_{\rme} = -1$,
$\ell_{Z} = \frac{e}{\sin \theta_{W}^0 \cos\theta_{W}^0 } [I_{\rme}^{3} - Q_{e}\sin^{2}\theta_{W}^0 ]$,
$r_{Z} = \frac{e}{\sin \theta_{W}^0 \cos\theta_{W}^0 } [- Q_{e}\sin^{2} \theta_{W}^0 ]$,
 $I^{3}_{\rme} = -\frac{1}{2}$ in the SM. 
Here $e$, $I_{\rme}^{3}$ and $\theta_{W}^0$ are the electromagnetic coupling, weak isospin of the electron 
and the bare Weinberg angle defined in (\ref{bareWangle}), respectively.

When  initial state electrons and positrons are longitudinally polarized, 
the differential cross section is given by
\begin{align}
\frac{d\sigma}{d\cos\theta}(P_{\rme^{-}}, P_{\rme^{+}})
&= \frac{1}{4} \biggl\{
(1+ P_{\rme^{-}}) (1+P_{\rme^{+}}) 
\frac{d\sigma_{\rme^{-}_{R} \rme^{+}_{R} }}{d\cos\theta}
+
(1 - P_{\rme^{-}}) (1 - P_{\rme^{+}}) 
\frac{d\sigma_{\rme^{-}_{L} \rme^{+}_{L} }}{d\cos\theta}
\nonumber\\& \qquad
+
(1 + P_{\rme^{-}}) (1 - P_{\rme^{+}}) 
\frac{d\sigma_{\rme^{-}_{R} \rme^{+}_{L} }}{d\cos\theta}
+
(1 - P_{\rme^{-}}) (1 + P_{\rme^{+}}) 
\frac{d\sigma_{\rme^{-}_{L} \rme^{+}_{R} }}{d\cos\theta}
\biggr\},
\end{align}
where $P_{\rme^{-}}$ and $P_{\rme^{+}}$ are the polarization of the electron and positron beam, respectively. $P_{\rme^{-}}=+1$ ($P_{\rme^{+}}=+1$) denotes purely right-handed electrons (positrons) \cite{MoortgatPick:2005cw,Arbuzov:2020ghr}.
$\sigma_{\rme_{X}^{-} \rme_{Y}^{+}}$ ($X,Y=L,R$) denotes the cross section 
for left-handed or right-handed electron and positron. 
When the electron mass is neglected, these cross sections 
are given by
\begin{align}
\frac{d\sigma_{\rme^-_L \rme^+_R}}{d\cos\theta}
&= \frac{1}{8\pi s}
\left(  u^2 |S_{LL} + T_{LL}|^2 + t^2 |S_{LR}|^2 \right),
\nonumber \\
\frac{d\sigma_{\rme^-_R \rme^+_L}}{d\cos\theta}
&= \frac{1}{8\pi s}
\left(  u^2 |S_{RR} + T_{RR}|^2 + t^2 |S_{LR}|^2 \right),
\nonumber \\
\frac{d\sigma_{\rme^-_L \rme^+_L}}{d\cos\theta}
&= \frac{d\sigma_{\rme^-_R \rme^+_R}}{d\cos\theta}
= \frac{1}{8\pi s} \cdot \left(s^2 |T_{LR}|^2 \right).
\end{align}

When $s,t \ll M_{Z}^{2}$, 
the cross section is  approximated by the one at the QED level,
where we obtain $S_{LL} = S_{RR} = S_{LR} = e^2/s$ and
$T_{LL} = T_{RR} = T_{LR} = e^2/t$, and 
\begin{align}
\frac{d\sigma_{\rm QED}}{d\cos\theta}(P_{e^-},P_{e^+})
&= 
\frac{e^{4}}{16\pi s}\left\{
(1 - P_{\rme^{-}}P_{\rme^{+}} ) \frac{t^{4}+u^{4}}{s^{2}t^{2}}
+ (1 + P_{\rme^{-}} P_{\rme^{+}}) \frac{s^{2}}{t^{2}}
\right\}.
\end{align}
For unpolarized electron or positron beams, the above expression reduces to
a familiar formula
\begin{align}
\frac{d\sigma^{\rm unpolarized}_{\rm QED}}{ d\cos\theta }
&=
 \frac{e^4}{16\pi s} \frac{s^4 + t^4 + u^4}{s^2 t^2}.
\end{align}
We also note that in terms of building blocks \eqref{eq:blocks} we 
can write down components of s-, t-channels, and interference terms as
\begin{align}
\frac{d\sigma}{d\cos\theta}
&=\frac{d\sigma^{\text{s-channel}}}{d\cos\theta}
+ \frac{d\sigma^{\text{t-channel}}}{d\cos\theta}
+ \frac{d\sigma^{\text{interference}}}{d\cos\theta},
\end{align}
where each component is given by
\begin{align}
\frac{d\sigma^{\text{s-channel}}}{d\cos\theta}(P_{\rme^{-}},P_{\rme^{+}})
&= \frac{1}{32\pi s} \biggl\{
  (1+P_{\rme^{-}})(1-P_{\rme^{+}}) \left[ u^{2} |S_{RR}|^{2} + t^{2} |S_{LR}|^{2} \right]
\nonumber\\& \qquad
+ (1-P_{\rme^{-}})(1+P_{\rme^{+}}) \left[ u^{2} |S_{LL}|^{2} + t^{2} |S_{LR}|^{2} \right]
\biggr\},
\nonumber \\
\frac{d\sigma^{\text{t-channel}}}{d\cos\theta}(P_{\rme^{-}},P_{\rme^{+}})
&= \frac{1}{32\pi s}\biggl\{
  (1+P_{\rme^{-}})(1+P_{\rme^{+}}) s^{2} |T_{LR}|^{2}
+ (1-P_{\rme^{-}})(1-P_{\rme^{+}}) s^{2} |T_{LR}|^{2}
\nonumber\\& \qquad
+ (1+P_{\rme^{-}})(1-P_{\rme^{+}}) u^{2} |T_{RR}|^{2}
+ (1-P_{\rme^{-}})(1+P_{\rme^{+}}) u^{2} |T_{LL}|^{2}
\biggr\},
\nonumber \\
\frac{d\sigma^{\text{interference}}}{d\cos\theta}(P_{\rme^{-}},P_{\rme^{+}})
&= \frac{1}{16\pi s} u^{2} \biggl\{
 (1+P_{\rme^{-}})(1-P_{\rme^{+}}) \Re(S_{RR} T_{RR}^{*})
\nonumber\\& \qquad
+(1-P_{\rme^{-}})(1+P_{\rme^{+}}) \Re(S_{LL} T_{LL}^{*})
\biggr\}.
\end{align}

When initial electrons and/or positrons are longitudinally polarized, one can measure left-right asymmetries.
The left-right asymmetry of polarized cross sections is given by
\begin{align}
A_{\LR}(P_{-},P_{+})
&\equiv \frac{
\sigma(P_{\rme^{-}} = -P_{-}, P_{\rme^{+}}= - P_{+})
- \sigma(P_{\rme^{-}} = +P_{-}, P_{\rme^{+}}= + P_{+})
}{
\sigma(P_{\rme^{-}} = -P_{-}, P_{\rme^{+}}= - P_{+})
+ \sigma(P_{\rme^{-}} = +P_{-}, P_{\rme^{+}}= + P_{+})
}
\nonumber \\
&= (P_{-} -  P_{+}) \cdot 
\frac{
\sigma_{\rme^{-}_{L} \rme^{+}_{R}} - \sigma_{\rme^{-}_{R} \rme^{+}_{L}}
}{
(1 + P_{-} P_{+}) (\sigma_{\rme^{-}_{L} \rme^{+}_{L}} + \sigma_{\rme^{-}_{R} \rme^{+}_{R}})
+
(1 - P_{-} P_{+}) (\sigma_{\rme^{-}_{L} \rme^{+}_{R}} + \sigma_{\rme^{-}_{R} \rme^{+}_{L}})
}
,
\nonumber \\ 
& 1\ge P_{-} \ge 0,\quad 1 \ge P_{+} \ge -1,
\label{eq:ALR}
\end{align}
where the cross section in a given bin $[\theta_{1},\theta_{2}]$ is given by
$
\sigma \equiv \int_{\cos\theta_{1}}^{\cos\theta_{2}} \frac{d\sigma}{d\cos\theta} d\cos\theta
$.
We have used $\sigma_{\rme^{-}_{L} \rme^{+}_{L}} = \sigma_{\rme^{-}_{R} \rme^{+}_{R}}$
because $\frac{d\sigma_{\rme^{-}_{L} \rme^{+}_{L}}}{d\cos\theta} - 
 \frac{d\sigma_{\rme^{-}_{R} \rme^{+}_{R}}}{d\cos\theta} = 0$.
We can also define the left-right asymmetry of the differential cross section as
\begin{align}
\lefteqn{A_{\LR}(P_{-},P_{+},\cos\theta)}
\quad
\nonumber\\
&\equiv
\frac{
\dfrac{d\sigma}{d\cos\theta}(P_{\rme^{-}} = -P_{-}, P_{\rme^{+}}= - P_{+})
- \dfrac{d\sigma}{d\cos\theta}(P_{\rme^{-}} = +P_{-}, P_{\rme^{+}}= + P_{+})
}{
\dfrac{d\sigma}{d\cos\theta}(P_{\rme^{-}} = -P_{-}, P_{\rme^{+}}= - P_{+})
+ \dfrac{d\sigma}{d\cos\theta}(P_{\rme^{-}} = +P_{-}, P_{\rme^{+}}= + P_{+})
}
\nonumber \\
&= \frac{
(P_{-} +  P_{+}) \left(\dfrac{d\sigma_{\rme^{-}_{L} \rme^{+}_{L}}}{d\cos\theta} 
- \dfrac{d\sigma_{\rme^{-}_{R} \rme^{+}_{R}}}{d\cos\theta} \right)
+
(P_{-} -  P_{+}) \left(\dfrac{d\sigma_{\rme^{-}_{L} \rme^{+}_{R}}}{d\cos\theta}
 - \dfrac{d\sigma_{\rme^{-}_{R} \rme^{+}_{L}}}{d\cos\theta} \right)
}{
(1 + P_{-} P_{+}) \left(\dfrac{d\sigma_{\rme^{-}_{L} \rme^{+}_{L}}}{d\cos\theta}
 + \dfrac{d\sigma_{\rme^{-}_{R} \rme^{+}_{R}}}{d\cos\theta} \right)
+
(1 - P_{-} P_{+}) \left(\dfrac{d\sigma_{\rme^{-}_{L} \rme^{+}_{R}}}{d\cos\theta}
 + \dfrac{d\sigma_{\rme^{-}_{R} \rme^{+}_{L}}}{d\cos\theta} \right)
}
\nonumber \\
&= (P_{-} - P_{+}) \cdot
\frac{
 \Sigma_{LR-RL}
}{(1 + P_{-} P_{+}) \Sigma_{LL+RR} + (1-P_{-} P_{+}) \Sigma_{LR+RL}},
\end{align}
where we have used $d\sigma_{\rme^{-}_{L} \rme^{+}_{L}}/d\cos\theta =d\sigma_{\rme^{-}_{R} \rme^{+}_{R}}/d\cos\theta  $ and defined
\begin{align}
\Sigma_{LL+RR} &\equiv 2s^{2} |T_{LR}|^{2},
\nonumber \\
\Sigma_{LR+RL} &\equiv u^{2} (|S_{LL} + T_{LL}|^{2} + |S_{RR} + T_{RR}|^{2}) + 2t^{2} |S_{LR}|^{2},
\nonumber \\
\Sigma_{LR-RL} &\equiv  u^{2} ( |S_{LL} + T_{LL}|^{2} - |S_{RR} + T_{RR}|^{2}  ).
\end{align}

In linear colliders we can measure the cross sections 
for $(P_{\rme^{-}},P_{\rme^{+}}) = (P_{-},P_{+})$, $(P_{-},-P_{+})$, $(-P_{-},P_{+})$
and $(-P_{-},-P_{+})$.
Combining these quantities, one can make a new observable quantity
which does not depend on the value of $P_{\pm}$.
In $\rme^+ \rme^- \to \rme^+ \rme^-$ scatterings we have
$A_{LR}(P_{-},+P_{+},\cos\theta)$ and $A_{LR}(P_{-},-P_{+},\cos\theta)$ as independent observables
and one may define the following non-trivial quantity:
\begin{align}
&A_{X}(\cos\theta) \equiv 
\frac{\Sigma_{LL+RR} - \Sigma_{LR+RL}}{\Sigma_{LL+RR} + \Sigma_{LR+RL}}
\nonumber \\
&\quad
= \frac{1}{P_{-} P_{+}} \cdot \dfrac{
 (P_{-} - P_{+}) A_{\LR}(P_{-},-P_{+},\cos\theta) - (P_{-} + P_{+}) A_{\LR}(P_{-}, +P_{+}\cos\theta)
}{
 (P_{-} - P_{+}) A_{\LR}(P_{-},-P_{+},\cos\theta) + (P_{-} + P_{+}) A_{\LR}(P_{-}, +P_{+},\cos\theta)
}.\label{eq:AX}
\end{align} 
where the second equality holds only when $P_{\pm}\ne0$ and $|P_{+}| \ne |P_{-}|$.
As is evident in the first line of \eqref{eq:AX}, $A_{X}(\cos\theta)$ is independent of 
the magnitudes of
polarization $P_{\pm}$.
This quantity may be used to explore NP beyond the SM as discussed below.

Since $\rme^+ \rme^- \to \rme^+ \rme^-$ scattering contains $t$-channel processes,
forward scatterings dominate. Therefore unlike the $\rme^+\rme^- \to f\bar{f}$ ($f\ne \rme^{-}$) scattering, 
the forward-backward asymmetry of $\rme^{+} \rme^{-} \to \rme^{+} \rme^{-} $ scattering is a less-meaningful quantity.

We note that all of the above formulas can be applied
to $\ell^{+}\ell^{-} \to \ell^{+}\ell^{-}$  ($\ell = \mu,\tau$) scatterings.

\section{Numerical Study}

In the followings we calculate $\rme^{+} \rme^{-} \to \rme^{+}\rme^{-}$
cross sections both in the SM and GHU models,
and we evaluate  effects of $Z'$ bosons in GHU models on observables
given in the previous section.
As benchmark points, we have chosen typical parameters of the A- and B-models
in Tables \ref{tbl:model} and \ref{tbl:couplings}. 
For experimental parameters we choose
$\sqrt{s}=250$ GeV and $\Lint = 250$ fb$^{-1}$ as typical value of linear $\rme^{+}\rme^{-}$ colliders like ILC\cite{ILC}. 
We also choose $\Lint = 2$ ab$^{-1}$, which will be achieved
at circular $\rme^{+}\rme^{-}$ colliders like FCC-ee\cite{Blondel:2021ema} and CEPC\cite{CEPCStudyGroup:2018ghi}. 
For the new asymmetry $A_{X}(\cos\theta)$ in \eqref{eq:AX},
 we consider $\sqrt{s} = 3$ TeV for future linear colliders like  CLIC\cite{CLICdesign}.
As for the longitudinal polarization, we set the parameter ranges
$- 0.8 \le P_{\rme^{-}} \le +0.8$ and $-0.3 \le P_{\rme^{+}} \le 0.3$,
which can be achieved at ILC\cite{ILC}.

\begin{figure}[H]
\centering
\includegraphics[width=0.47\textwidth]{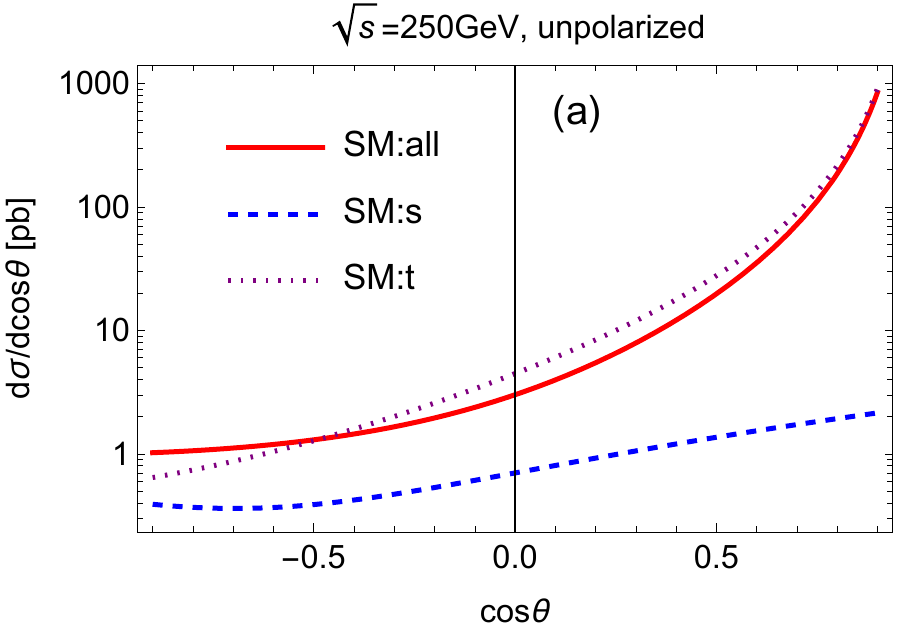}
\includegraphics[width=0.47\textwidth]{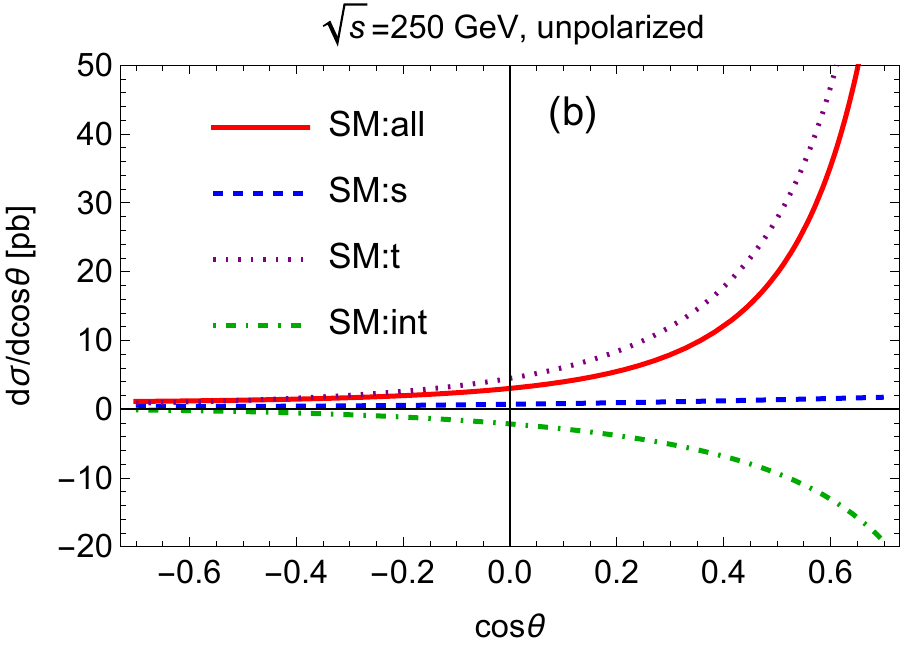}
\caption{%
Differential cross sections for unpolarized $\rme^{+}\rme^{-}$ initial
states in the SM.
(a) Log-scale plot with $-0.9 \le \cos\theta \le 0.9$.
(b) Linear plot with $-0.7 \le \cos\theta \le 0.7$.
In both plots, red-solid lines indicate the total of $s$-, $t$-channels and interferences. 
The $s$-channel and $t$-channel contributions are drawn as blue-dashed and purple-dotted lines, respectively. 
In (b), the negative contribution from interferences is plotted with the green-dashed line.
}\label{fig:dsigma}
\end{figure}

In Figure~\ref{fig:dsigma}   $\rme^{+}\rme^{-}\to \rme^{+}\rme^{-}$ differential cross sections in the SM are plotted.
In the forward-scattering region ($\cos\theta>0$),
 the magnitudes of cross sections of $t$-channel and the interference
parts are much larger than those of the $s$-channel part.

\begin{figure}[H]
\includegraphics[width=0.48\textwidth]{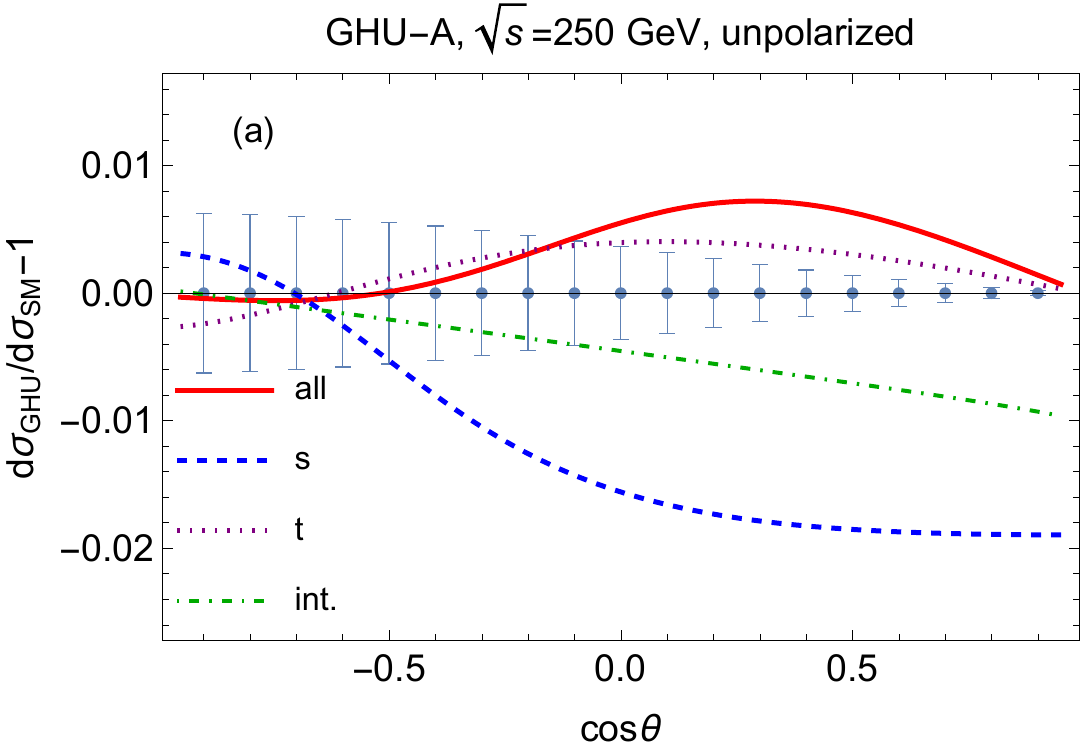}\hspace{0.2cm}
\includegraphics[width=0.48\textwidth]{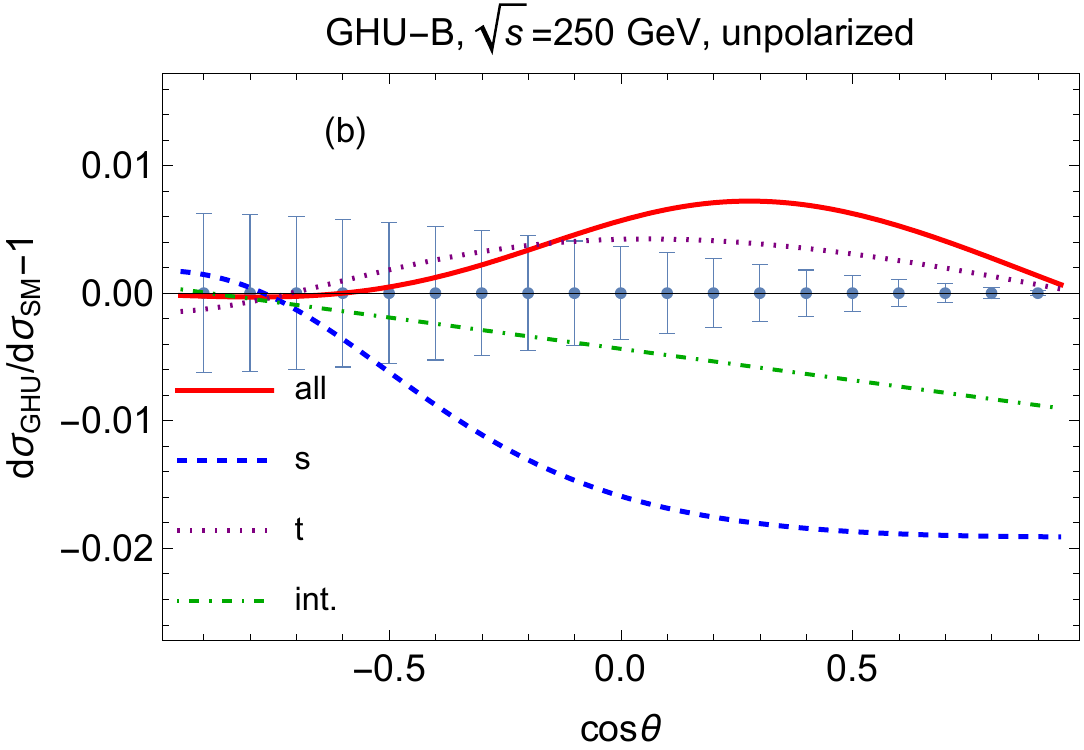}
\caption{
Deviations of differential cross sections 
of GHU from those in the SM,
$
\frac{d\sigma^{\GHU}}{d\cos\theta}/
\frac{d\sigma^{\SM}}{d\cos\theta}-1
$,
for unpolarized $\rme^{+}\rme^{-}$ beams
are plotted.  
The left plot is for the A-model and the right plot is for the B-model. In each plot,
the red-solid curve represents the deviation  of the sum of all the components of the differential cross section.
Blue-dashed, purple-dotted and green dot-dashed curves correspond to
 the deviations of $s$-, $t$-channels and interference components of differential cross sections, respectively.
 Error-bars are estimated for $\Lint = 250 \text{ fb}^{-1}$. 
}\label{fig:Delta-dsigma}
\end{figure}

\begin{figure}[H]
\centering
\includegraphics[width=0.48\textwidth]{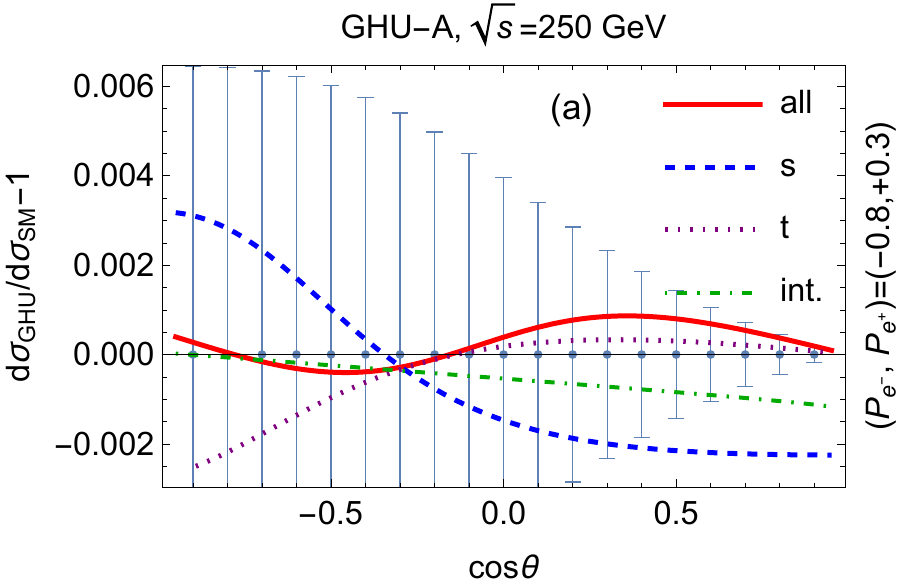}\hspace{0.2cm}
\includegraphics[width=0.48\textwidth]{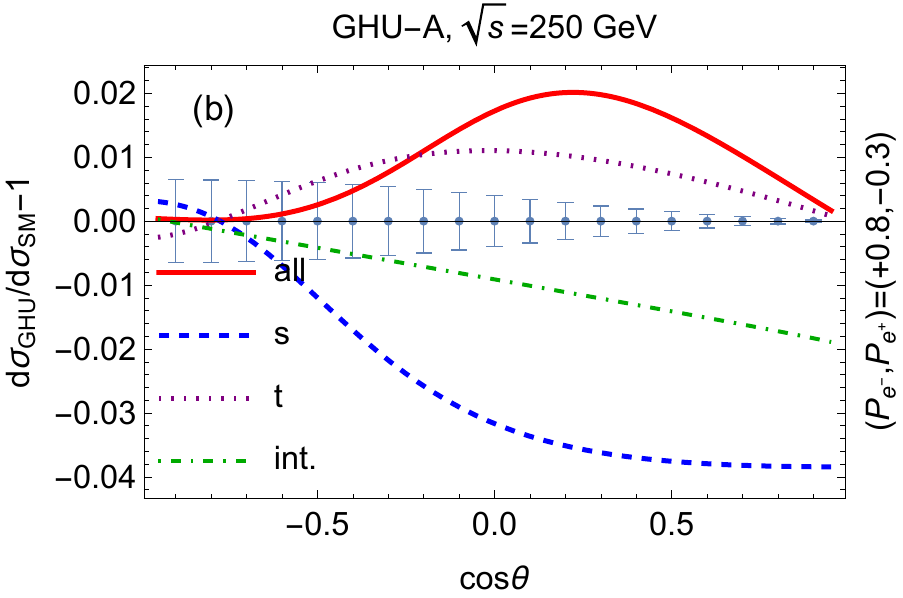}\\
\includegraphics[width=0.48\textwidth]{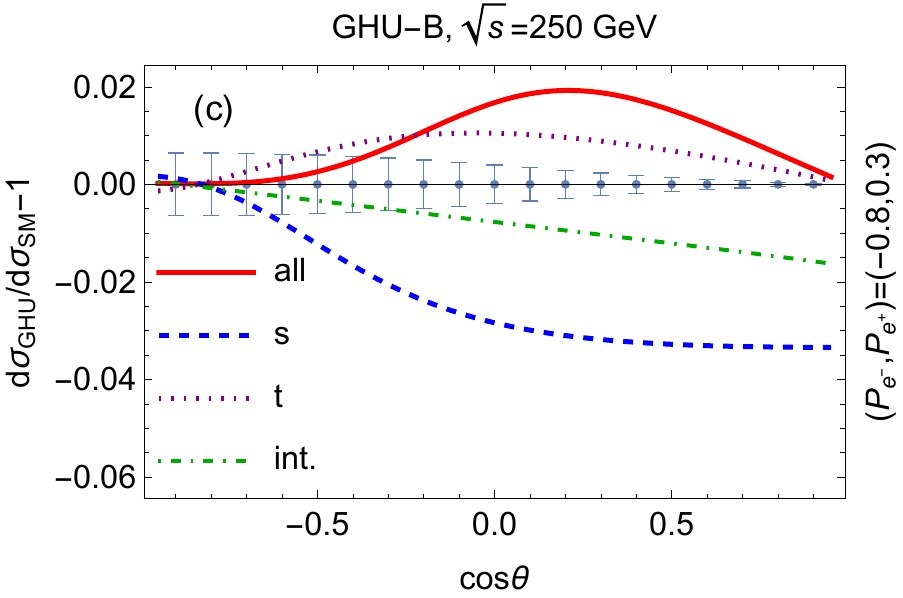}\hspace{0.2cm}
\includegraphics[width=0.48\textwidth]{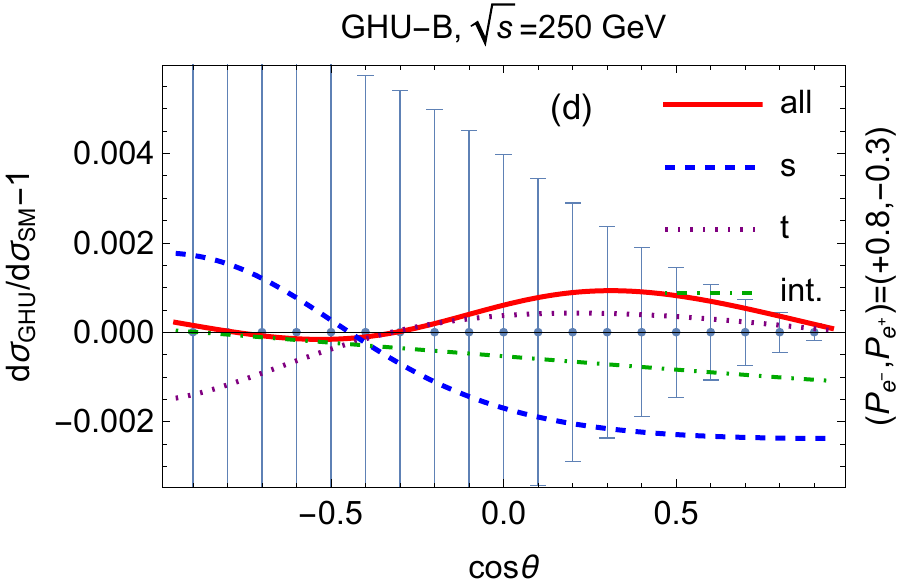}%
\caption{%
Deviations of differential cross sections
of GHU models from the SM 
for polarized $\rme^{+}\rme^{-}$ beams.
GHU-A [(a) and (b)] and GHU-B [(c) and (d)].
(a) and (c) are for $(P_{e^{-}},P_{e^{+}}) = (-0.8,+0.3)$.
(b) and (d) are for$(P_{e^{-}},P_{e^{+}}) = (+0.8,-0.3)$.
Meanings of the curves and error-bars are the same as in Figure~\ref{fig:Delta-dsigma}.
}\label{fig:Delta-dsigma0}
\end{figure}
In Figures~\ref{fig:Delta-dsigma} and \ref{fig:Delta-dsigma0}, 
the differences of differential cross sections of the GHU from the SM for unpolarized and polarized beams are plotted, respectively.
In the figures, differences of $s$-channel, $t$-channel and interference contributions are also plotted.
In the $s$-channel, the NP effects contribute destructively in the forward scattering.
On the other hand, in the $t$-channel NP effects contribute constructively.
Since the cross section is dominated by $t$-channel, 
the total of 
$s$-, $t$- and interference channels 
increases due to the NP effects.

In the A-model $Z'$ bosons have larger couplings to right-handed electrons than to left-handed electrons.
Therefore
the cross section of the $\rme_{R}^{-}\rme_{L}^{+}$ initial states  becomes larger than that of $\rme_{L}^{-}\rme_{R}^{+}$. 
On the other hand,
in the B-model $Z'$ bosons have larger couplings to left-handed electrons than to right-handed electrons,
and the cross section of $\rme_{L}^{-}\rme_{R}^{+}$ initial states becomes larger.

NP effects become smaller when $\theta$ becomes smaller.
The statistical uncertainty, however, also becomes small since the cross section
becomes very large.
Therefore deviations of the cross section relative to statistical uncertainties may become large.

For unpolarized $\rme^{+}\rme^{-}$ beams (Figure~\ref{fig:Delta-dsigma0}),
the new physics effect in  both models tends to enhance the cross section at
forward scattering with almost the same magnitude.
In the B-model the suppression of NP effects due to larger $Z'$ masses 
is compensated by larger couplings of $Z'$ bosons than the couplings  in the A-model.
The enhancement of the differential cross section due to the NP effects at $\cos\theta \sim 0.3$ is around $1\,\%$.

For polarized beams  deviations can be much larger.
In the A-model, electrons have large right-handed couplings to $Z'$ bosons and for right-handed polarized beam relative deviations of the cross-section 
become as much as $2\,\%$ [Figure~\ref{fig:Delta-dsigma0}-(b)], whereas for right-handed beams relative deviations are around $0.1\,\%$ [Figure~\ref{fig:Delta-dsigma0}-(a)].
Contrarily, in the B-model a left-handed electron has large couplings to $Z'$ bosons. Therefore in the B-model deviations become large for left-polarized beam [Figures~\ref{fig:Delta-dsigma0}-(c) and (d)].
In Figure \ref{fig:Delta-dsigma},
we have also shown statistical 1 $\sigma$ relative errors at $\Lint =250$ fb$^{-1}$ 
for bins
$[\cos\theta_{0}-0.05,\cos\theta_{0}+0.05]$ ($\cos\theta_{0} = -0.90, -0.80,\dots, +0.90$) as vertical bars.
In each bin, the observed number of events and statistical uncertainty are given by $N$ and $\sqrt{N}$, respectively.
Therefore relative error of the cross section is 
estimated as the inverse of the square of the number of events $N$.
\begin{figure}[H]
\centering
\includegraphics[width=0.58\textwidth]{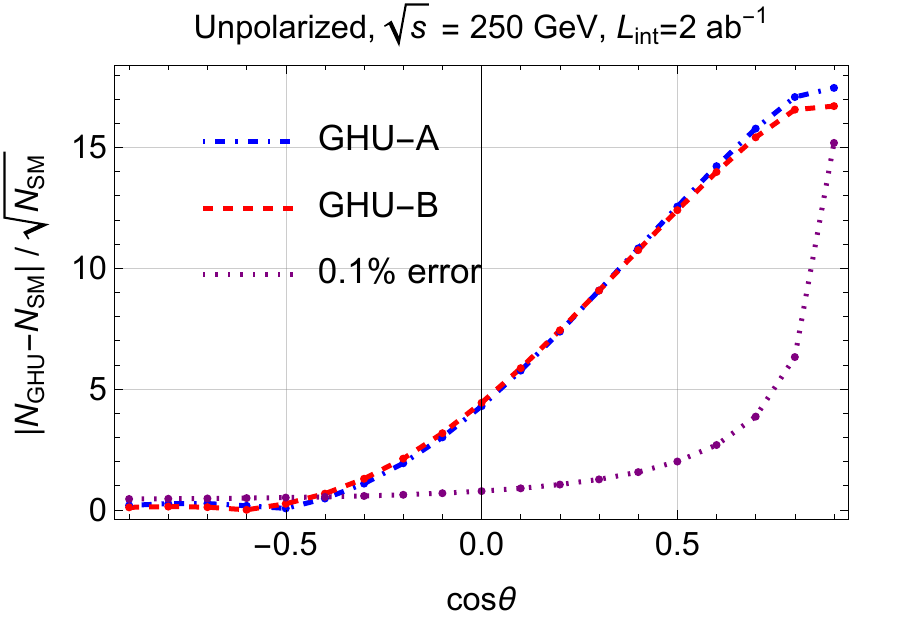}
\caption{Estimated significances on differential cross sections of GHU models
with unpolarized $\rme^{+} \rme^{-}$ beam with the integrated luminosity $\Lint = 2\text{ ab}^{-1}$.
A statistical significance of 0.1\,\% non-statistical error is plotted with a purple dot-dashed line.}
\label{fig:sigdsigma}
\end{figure}

In Figure~\ref{fig:sigdsigma}, the statistical significances in the GHU models are plotted.
An estimated significance of the deviation of the cross section 
in a bin is given by
\begin{align}
\frac{|N_{\GHU} - N_{\SM}|}{N_{\SM}} \bigg/ \frac{\sqrt{N_{\SM}}}{N_{\SM}}
= \frac{|N_{\GHU} - N_{\SM}|}{\sqrt{N_{\SM}}},
\end{align} 
where $N_{\GHU}$ and $N_{\SM}$ are observed numbers of events in a bin.
In Figure \ref{fig:sigdsigma},
significances are larger than 5 $\sigma$
for $\cos\theta\gtrsim0.1$. 
Significances are very large for forward scatterings,
but are very small for backward scatterings.
In Figure \ref{fig:sigdsigma}, relative 0.1\,\% errors are also shown. Errors due to the NP effects become very small and are around $0.1\,\%$ for $\cos\theta \simeq 0.9$. 
A similar analysis has been given in Ref.\cite{Richard:2018zhl}.

\begin{figure}[H]
\centering
\includegraphics[width=0.55\textwidth]{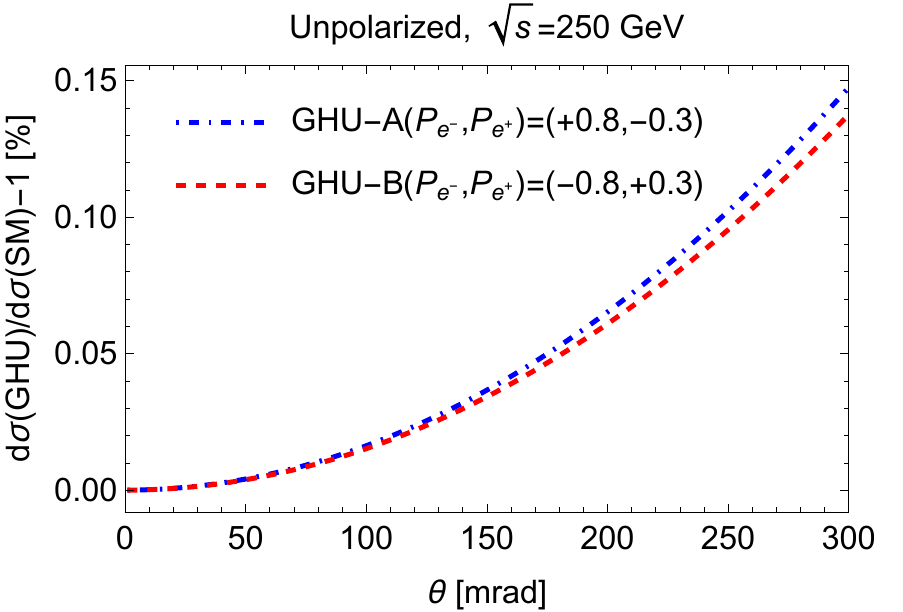}
\caption{Deviations of differential cross sections of GHU models from the SM in the forward-scattering region ($\theta \le 300$\text{ mrad}).}\label{fig:farforward}
\end{figure}

For small scattering angles $\theta$, the scattering amplitude is dominated by
$t$-channel contributions which are constructed with the blocks $T_{LL}$, $T_{RR}$ and $T_{LR}$.
When $|t| \simeq s\theta^{2}/4 \ll M_{Z}^{2}, M_{Z'}^{2}$,
we can approximate the SM and NP contributions to the block $T_{LL}$ in the scattering amplitude as 
\begin{align}
T_{LL}^{\NP} &\equiv 
\sum_{Z'} \frac{\ell_{Z'}^{2}}{t - M_{Z'}^{2} + iM_{Z'}\Gamma_{Z'}}
\simeq 
\sum_{Z'} \frac{\ell_{Z'}^{2} }{- M_{Z'}^{2}+ i M_{Z'}\Gamma_{Z'}}.
\end{align}
When $s\theta^{2}/4 \ll M_{Z}^{2}$, 
$T_{LL}$ is dominated by the QED part $T_{LL}^{\rm QED} \simeq -4e^{2}/(\theta^{2}s)$ and the NP effects are estimated as
\begin{align}
\frac{T_{LL}^{\NP}}{T_{LL}^{\rm QED}} &\simeq \theta^{2} \sum_{Z'} \frac{
(\ell_{Z'}^{2}/4e^{2})s }{ M_{Z'}^{2} - i M_{Z'}\Gamma_{Z'}}
= \theta^{2} \cdot \calO(s/M_{Z'}^{2}),
\end{align}
and similar analysis is applied to $T_{LR}$ and $T_{RR}$.
Consequently, this correction arises not only in amplitudes but also in
differential cross sections.
For $\sqrt{s} = 250\,\GeV$ and  $\theta \lesssim 300\text{ mrad}$,
the QED $t$-channel contribution dominates 
and corrections due to $Z'$ bosons are suppressed by a factor $\theta^{2}s/M_{Z'}^{2}$.
In Figure~\ref{fig:farforward}, deviations of differential cross sections of GHU from the SM for $\theta<300\text{ mrad}$ are plotted.
Deviations of cross sections from the SM are proportional to the square of 
$\theta$ and become smaller than $0.1\,\%$ when $\theta < 250\text{ mrad}$.
The measurement of Bhabha scatterings at small scattering angle is used for the determination of the luminosity of $\rme^{+} \rme^{-}$ collision and uncertainties of the luminosity should be smaller than $0.1\,\%$. In GHU models the NP effects on such uncertainty are well suppressed when $\theta \lesssim 100\text{ mrad}$.
At ILC, the luminosity calorimeter in ILD (SiD) operates between 43 and 68 (40 and 90) mrad \cite{ILC}, where the influence of the $Z'$ bosons is below $0.1\,\%$.

\begin{figure}[H]
\centering
\includegraphics[width=0.48\textwidth]{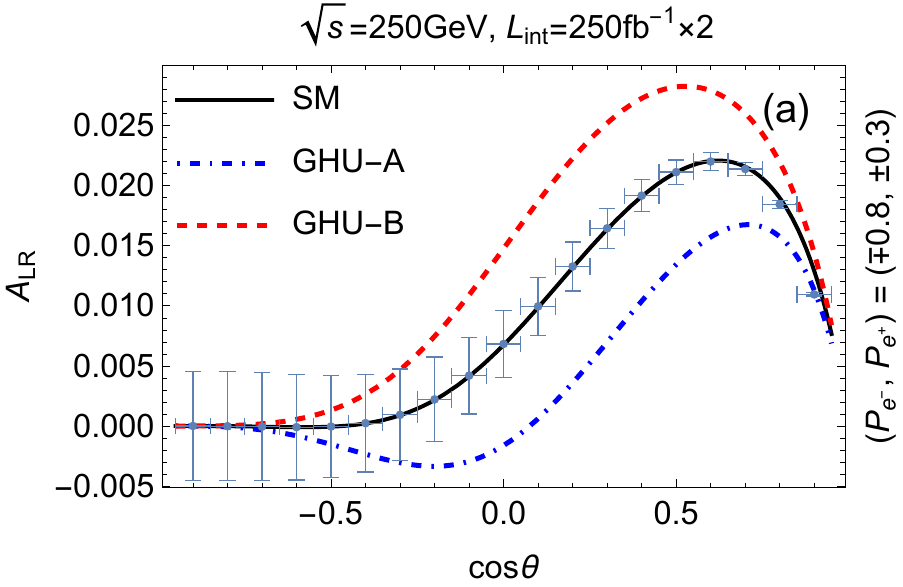}
\includegraphics[width=0.48\textwidth]{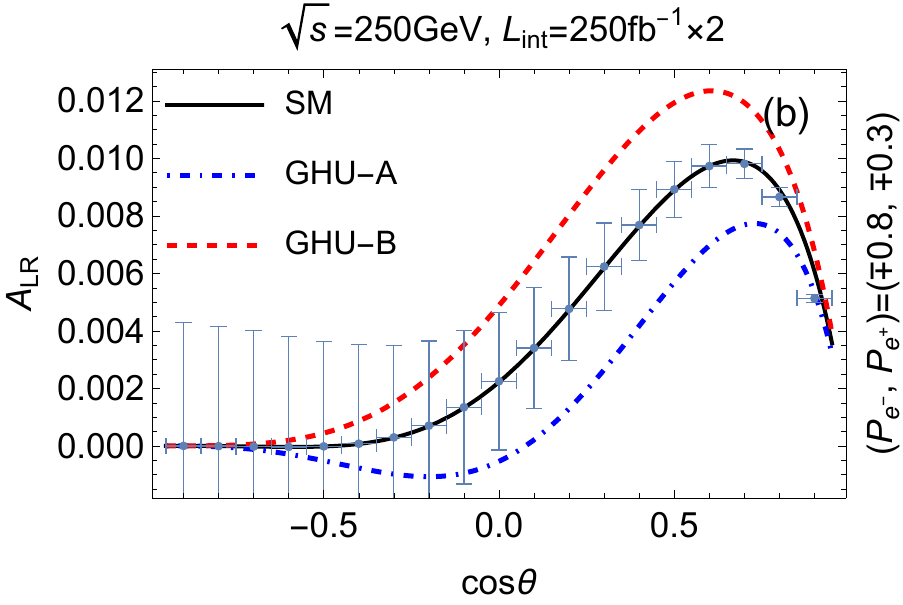}
\caption{Left-right asymmetries. In both plots,
blue-dotdashed, and red-dashed and black-solid curves correspond to
 the GHU-A, GHU-B and the SM, respectively. 
 Error-bars are indicated for $\Lint = 250\,\fbi$ in each polarization. 
 (a) Asymmetries for $(P_{\rme^{-}},P_{\rme^{+}}) = (\mp0.8,\pm0.3)$.
 (b) Asymmetries for $(P_{\rme^{-}},P_{\rme^{+}}) = (\mp0.8,\mp0.3)$. }\label{fig:ALR}
\end{figure}

When the initial electron and positron beams are longitudinally polarized,
the left-right asymmetry $A_{\LR}$ \eqref{eq:ALR} can be measured.
In Figure~\ref{fig:ALR}, the left-right asymmetries of the SM and GHU models are plotted.
The measured asymmetries become larger when $|P_{\rme^{-}} - P_{\rme^{+}}|$ are larger. 
As seen in Figure~\ref{fig:Delta-dsigma}, in the A-model cross section of $\rme_{R}^{-}\rme_{L}^{+}$ initial states becomes large whereas
in the B-model cross section of $\rme_{L}^{-}\rme_{R}^{+}$ initial states is enhanced
due to the large left-handed $Z'$ couplings.
Therefore $A_{\LR}$ of B-models are larger than the SM, whereas $A_{\LR}$ of A-models are smaller. 
 Since the $A_{\LR}$ is proportional to $|P_{\rme^{-}} - P_{\rme^{+}}|$,
the asymmetries in Figure~\ref{fig:ALR}-(a) are almost twice as large as in (b).

In Figure \ref{fig:ALR}, an asymmetry $A_{\LR}$ in a bin and the 
statistic error $\Delta A_{\LR}$ are also shown. 
Here
\begin{align}
A_{\LR} &= \frac{N_{L} - N_{R}}{N_{L}+N_{R}},
&
\Delta A_{\LR} &= 
\sqrt{\frac{2 (N_{L}^{2} + N_{R}^{2})}{ (N_{L} + N_{R})^{3}}} 
\end{align} 
with $N_{L}$ and $N_{R}$ being observed number of events for the left-handed  ($P_{\rme^{-}}<0$) and right-handed ($P_{\rme^{-}}>0$) electron beams, respectively.
For small scattering angle $\cos\theta \gtrsim 0.8$, both 
$A_{\LR}^{\GHU}$ and $A_{\LR}^{\SM}$ 
become close to each other.

\begin{figure}[H]
\centering
\includegraphics[width=0.58\textwidth]{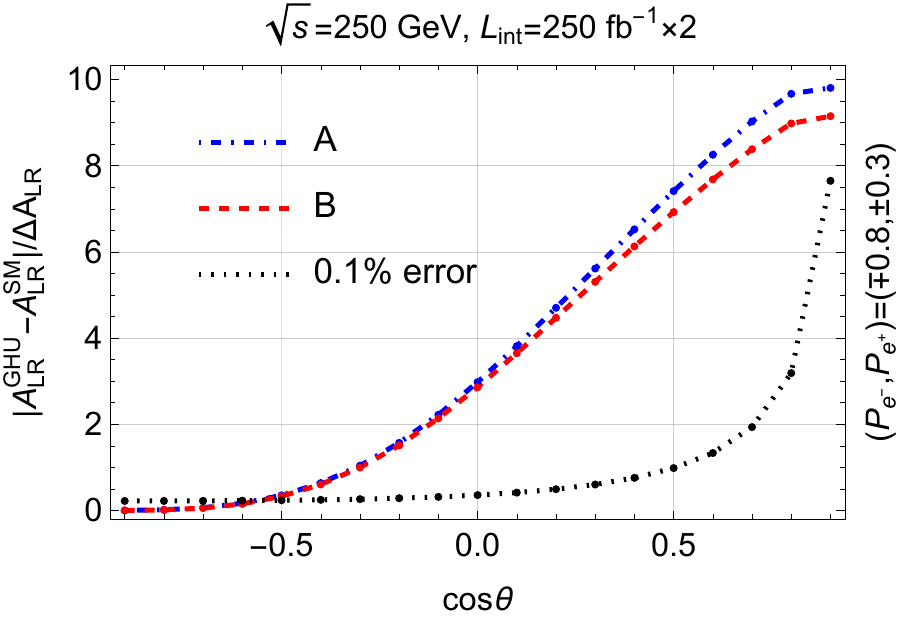}
\caption{%
Estimated statistical significance on left-right asymmetries in GHU models.
The blue-dotdashed, red-dashed and
black-dotted lines indicate
GHU-A, GHU-B and
 0.1\,\% non-statistical errors, respectively.}\label{fig:sigALR}
\end{figure}

To see how NP effects against statistical uncertainty grow for small $\theta$,
we plotted in Figure~\ref{fig:sigALR} the averages and statistical significances of
left-right asymmetries in GHU models in each bin which are estimated as  
\begin{align}
\frac{A_{LR}^{\GHU} - A_{LR}^{\SM} }{\Delta A_{\LR}}.
\end{align}
For the forward scattering with $\cos\theta \gtrsim 0.2$,
the deviations are bigger than several times of standard deviations.
In Figure~\ref{fig:sigALR}, 
the significance is larger than $5\,\sigma$ for $\cos\theta\gtrsim0.2$. 
Both models are well distinguished from the SM.
Using the magnitude and sign of deviations, the A-model
and B-model can be distinguished.

\begin{figure}[H]
\centering
\includegraphics[width=0.48\textwidth]{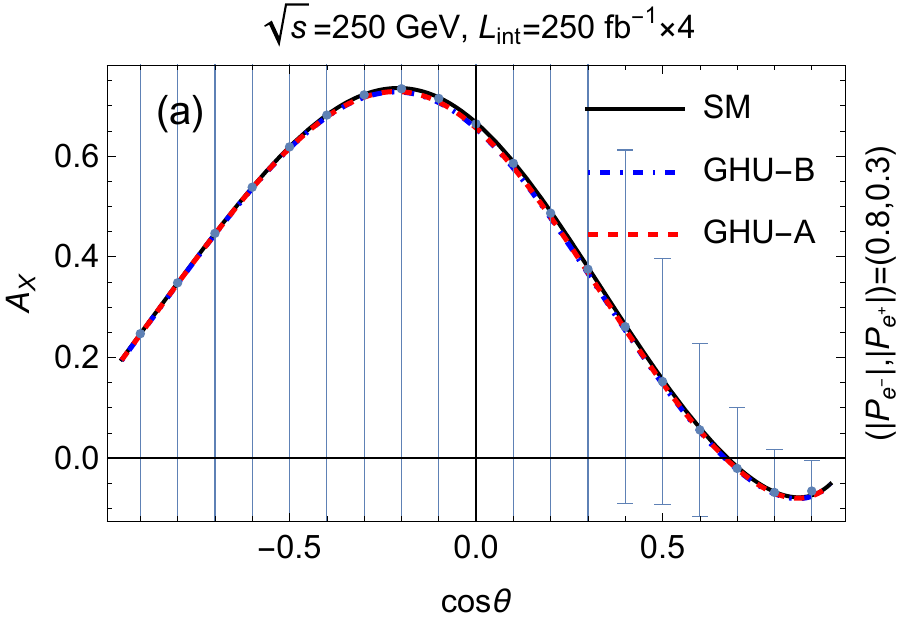}\hspace{0.2cm}
\includegraphics[width=0.48\textwidth]{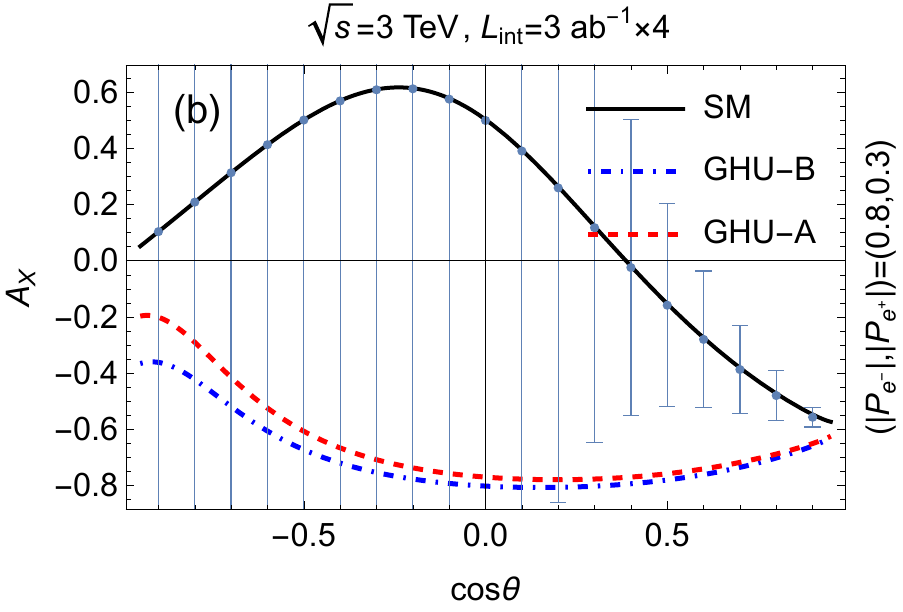}
\caption{Plot of an asymmetry $A_{X}$ in Eq.~\eqref{eq:AX}.
The left plot (a) is for $\sqrt{s} = 250\,\GeV$ with $\Lint = 250\,\fbi$ for each set of ($P_{\rme^{-}}, P_{\rme^{+}}$).
The right plot (b) is for $\sqrt{s} = 3\,\TeV$ with $\Lint = 3\,\abi$ for each set of ($P_{\rme^{-}}, P_{\rme^{+}}$).
The red-dashed, blue-dotdashed and black-solid curves correspond to the A-model, B-model and the SM, respectively.
 Error-bars are indicated for $\Lint = 250\,\fbi$ in (a) and
 $\Lint = 3\,\abi$ in (b) in each set of ($P_{\rme^{-}}, P_{\rme^{+}}$).}\label{fig:AX}
\end{figure}

In Figure~\ref{fig:AX},
the asymmetry $A_{X}$ defined in Eq.~\eqref{eq:AX} is plotted
for $\sqrt{s} = 250\,\GeV$ and 
$\sqrt{s} = 3\,\TeV$.
At $\sqrt{s} = 250\,\GeV$, the NP effect on $A_{X}$ is very small.
For $\sqrt{s} = 3\,\TeV$, the asymmetry $A_{X}$ of the SM and GHU models is clearly different and may be discriminated experimentally.
In the present analysis of NP effects, only first KK excited states of neutral bosons are taken into account. At $\sqrt{s} \sim  3\,\TeV$, effects of second KK modes on $A_{X}$ are estimated as a few percent. These effects are much smaller than the effects of the first KK modes.

\section{Summary}
In this paper we examined the effects of $Z'$ bosons
in the gauge-Higgs unification (GHU) models in the $\rme^{+}\rme^{-}\to \rme^{+}\rme^{-}$ (Bhabha) scatterings.
We first formulated  differential cross sections in Bhabha scattering including  $Z'$ bosons.
We then numerically evaluated the deviations of differential cross sections in the two 
$\SO(5) \times \U(1) \times \SU(3)$ GHU models (the A- and B-models) at $\sqrt{s} = 250\,\GeV$. 
We found that at $\Lint = 2\,\abi$ with unpolarized $\rme^{+}\rme^{-}$ beams, the deviation due to $Z'$ bosons 
in the GHU models from the SM can be clearly seen.
We also found that for $80\,\%$-longitudinally polarized electron and $30\,\%$-polarized positron beams, 
deviations of the differential cross sections from the SM 
become as large as a few percent for $\cos\theta \sim 0.2$, and that
the A-model and the B-model are well distinguished
at more than $3\,\sigma$ significance
at $\Lint =250\,\fbi$.
We also checked the effects of $Z'$ bosons are negligible
for the scattering angle smaller than $100$ mrad at $\sqrt{s} = 250\,\GeV$. Therefore Bhabha scatterings at very small $\theta$ 
can be safely used as the measurements of the luminosity in the $\rme^{+}\rme^{-}$ collisions.  
Finally we introduced the new observable $A_{X}$. 
We numerically evaluated it at $\sqrt{s} = 250\,\GeV$ and
$3\,\TeV$. 
Effects of the GHU models on $A_{X}$
can be seen at future TeV-scale $\rme^{+}\rme^{-}$ colliders.

In this paper the effects of $Z'$ bosons are calculated at the Born level. Higher-order QED effects should be taken 
into account for more precise evaluation \cite{Bardin:1990,Bardin:2017}.

\section*{Acknowledgements}

This work was supported in part 
by European Regional Development Fund-Project Engineering Applications of 
Microworld Physics (No.\ CZ.02.1.01/0.0/0.0/16\_019/0000766) (Y.O.), 
by the National Natural Science Foundation of China (Grant Nos.~11775092, 
11675061, 11521064, 11435003 and 11947213) (S.F.), 
by the International Postdoctoral Exchange Fellowship Program (IPEFP) (S.F.), 
and by Japan Society for the Promotion of Science, 
Grants-in-Aid  for Scientific Research, Nos.
JP19K03873 (Y.H.) and JP18H05543 (N.Y.).


\vskip 1.cm

\def\jnl#1#2#3#4{{#1}{\bf #2},  #3 (#4)}

\def\Zphys{{ Z.\ Phys.} }
\def\jssc{{ J.\ Solid State Chem.\ }}
\def\jpsJ{{ J.\ Phys.\ Soc.\ Japan }}
\def\ptps{{ Prog.\ Theoret.\ Phys.\ Suppl.\ }}
\def\PTP{{ Prog.\ Theoret.\ Phys.\  }}
\def\PTEP{{ Prog.\ Theoret.\ Exp.\  Phys.\  }}
\def\JMP{{ J. Math.\ Phys.} }
\def\NPB{ Nucl.\ Phys. \textbf{B}}
\def\NP{{ Nucl.\ Phys.} }
\def\PLB{{ Phys.\ Lett.} \textbf{B}}
\def\PL{{ Phys.\ Lett.} }
\def\PRL{Phys.\ Rev.\ Lett. }
\def\PRB{Phys.\ Rev. \textbf{B}}
\def\PRD{Phys.\ Rev. \textbf{D}}
\def\PRe{Phys.\ Rep. }
\def\AP{{Ann.\ Phys.\ (N.Y.)}}
\def\RMP{Rev.\ Mod.\ Phys. }
\def\ZPC{Z.\ Phys. \textbf{C}}
\def\SCI{Science}
\def\CMP{Comm.\ Math.\ Phys. }
\def\MPLA{Mod.\ Phys.\ Lett. \textbf{A}}
\def\EPJC{Eur.\ Phys.\ J. \textbf{C}}
\def\JHEP{JHEP }
\def\ibid{{\it ibid.} }


\renewenvironment{thebibliography}[1]
         {\begin{list}{[$\,$\arabic{enumi}$\,$]}  
         {\usecounter{enumi}\setlength{\parsep}{0pt}
          \setlength{\itemsep}{0pt}  \renewcommand{\baselinestretch}{1.2}
          \settowidth
         {\labelwidth}{#1 ~ ~}\sloppy}}{\end{list}}

 \end{document}